\pdfoutput=1

\documentclass{article}

\usepackage{arxiv}

\usepackage[utf8]{inputenc} 
\usepackage[T1]{fontenc}    
\usepackage{hyperref}       
\usepackage{url}            
\usepackage{booktabs}       
\usepackage{amsfonts}       
\usepackage{nicefrac}       
\usepackage{microtype}      

\usepackage{lipsum}         
\usepackage{graphicx}
\usepackage{natbib}
\usepackage{doi}
\usepackage{color}
\usepackage{amsbsy}

\usepackage{graphicx}
\usepackage{epstopdf}
\usepackage{subfigure}

\usepackage{algorithm}
\usepackage{algorithmicx}
\usepackage{algpseudocode}
\usepackage{amsmath}
\usepackage{array}
\usepackage{cleveref}       

\title{4D Agnostic Real-Time Facial Animation Pipeline for Desktop Scenarios}

\date{Mar. 15, 2023}

\author{ {Wei Chen} \\
	FACEGOOD \\
	\And
	{HongWei Xu}\thanks{Corresponding author} \\
	FACEGOOD \\
	\And
    {Jelo Wang} \\
    FACEGOOD \\
}

\renewcommand{\undertitle}{}

\hypersetup{
pdfauthor={Hongwei Xu},
pdfkeywords={Facial Capture, 3DMM, Non-rigid ICP},
}

\begin{document}
\maketitle

\begin{abstract}

We present a high-precision real-time facial animation pipeline suitable for animators to use on their desktops. This pipeline is about to be launched in FACEGOOD's Avatary\footnote{https://www.avatary.com/} software, which will accelerate animators' productivity. The pipeline differs from professional head-mounted facial capture solutions in that it only requires the use of a consumer-grade 3D camera on the desk to achieve high-precision real-time facial capture. The system enables animators to create high-quality facial animations with ease and speed, while reducing the cost and complexity of traditional facial capture solutions. Our approach has the potential to revolutionize the way facial animation is done in the entertainment industry.

\end{abstract}


\section{Introduction}

Facial capture refers to the process of capturing and digitizing the movement of a person's face using specialized technology. This technology enables the creation of high-fidelity digital replicas of human facial expressions and movements, which can be used in a wide range of applications such as film and video game production, virtual reality experiences, and teleconferencing.

Real-time facial animation is a challenging task in computer graphics, as it requires accurate and efficient processing of various facial features, such as the mouth, eyes, and eyebrows. In this context, the accurate calculation of weights and the precise tracking of facial landmarks and eyeball movement are essential for creating a natural and expressive facial animation\cite{xu2021high}. In this text, we will explore the methods and techniques used in real-time facial animation, including the calculation of weights, weights filter, and eye gaze.

\section{Face Reconstruction}
Our program uses a three-step process to reconstruct a face, which involves Fusion, 3DMM, and Non-rigid Iterative Closest Point (Non-rigid ICP). Firstly, Fusion technique is applied to merge multiple face images captured from different angles. Next, the 3DMM method is utilized to obtain a more realistic and precise 3D model of the face, which accurately captures the unique characteristics of human facial features. Finally, Non-rigid ICP is employed to refine the model by adjusting the position and orientation of the facial features, ensuring a seamless fit with the original images. The proper implementation of the 3DMM method is critical in creating a natural and lifelike face, with enhanced features that closely resemble those of actual people.
\begin{figure}[t]
	\centering
	\centering{\includegraphics[width=40pc]{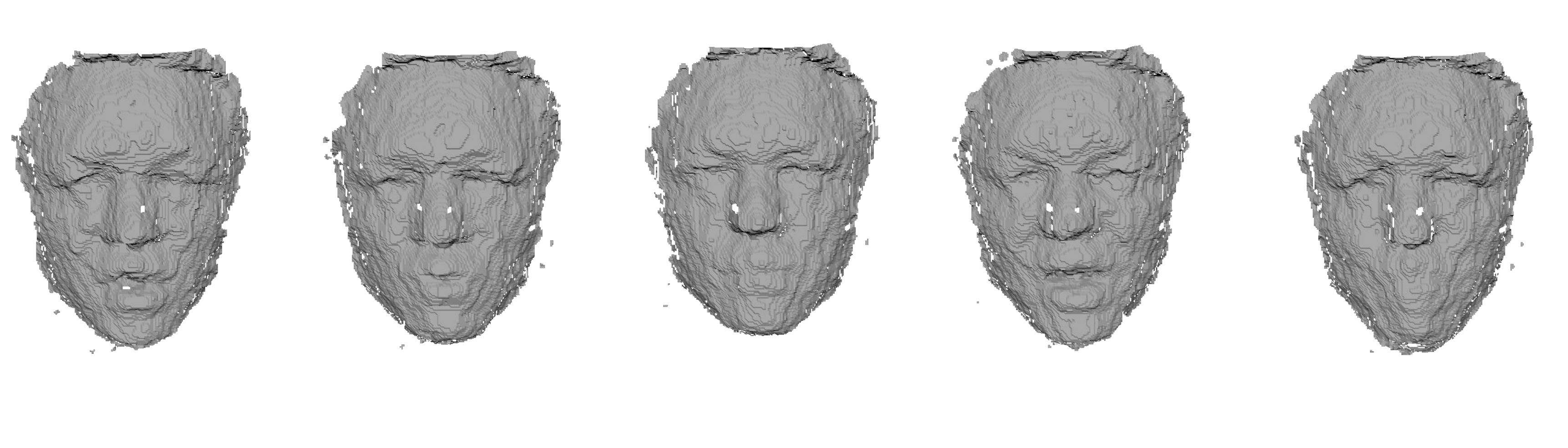}}
	\caption{The points cloud data for a single view.}
	\label{fig:1}
\end{figure}

\subsection{Fusion}
\begin{figure}[t]
	\centering{\includegraphics[width=40pc]{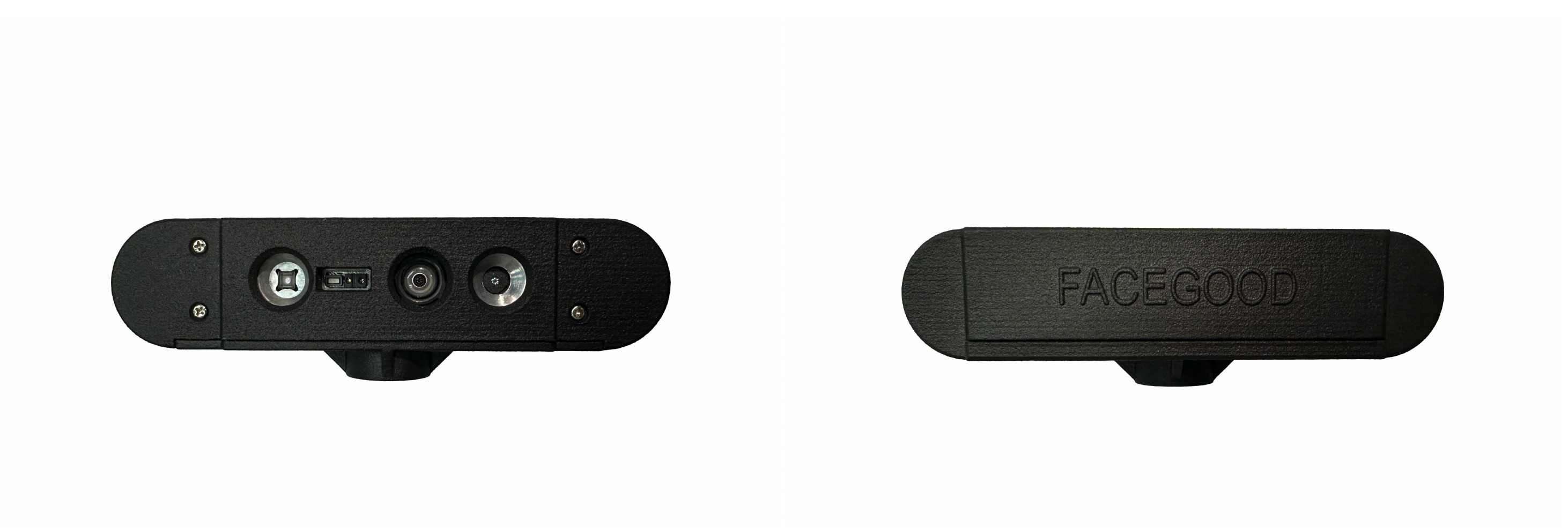}}
	\centering
	\caption{camera}
	\label{fig:2}
\end{figure}

Before performing fusion, we obtain multiple views of the RGB and depth maps(see Fig \ref{fig:1}) from a depth camera. Specifically, the depth camera we used is showed in Fig \ref{fig:2}.

The fusion method used is the kinect fusion\cite{izadi2011kinect}.We show some result of our fusion of different expressions in Fig \ref{fig:3}.

\subsection{3DMM}

To implement the 3DMM method, we utilized 400 manually scanned and topologized heads. A novel head model can be represented as a linear combination of the eigenvectors $\pmb{s}_i$ of the covariance matrix\cite{Blanz1999}. Consequently, it is necessary to compute the weights $\pmb{\omega}_{3DMM}$ of the basis vectors.

Initially, we start by obtaining the closest point $\pmb{p}_i$ in the scanned head for every point $\pmb{q}_i$ in the current template (which is initially the average head), and the normal vector of $\pmb{p}_i$ is represented as $\pmb{n}_i$. The discrepancy between the correspondences is calculated by utilizing the point-plane distance and point-point distance:
\begin{equation}\label{eq1}
 E_{3DMM-plane} = \sum_{i = 1}^{k} ||\pmb{n}_i^T(\pmb{p}_i - \pmb{q}_i')||_2^2 + 0.1||\pmb{p}_i - \pmb{q}_i'||_2^2.
 \end{equation}

The point $\pmb{q}_i'$ can be expressed as a linear combination of the base vectors and their corresponding weights in the following manner:
\begin{equation}\label{eq2}
\pmb{q}_i' = \bar{\pmb{s}}(i) + \sum_{j=1}^{m} \pmb{\omega}_{3DMM}^j \pmb{s}_j(i),
 \end{equation}
where $\bar{\pmb{s}}$ is the average shape, $\pmb{\omega}_{3DMM}^j$ is the weight of the $j$-th base, and $\pmb{s}_j(i)$ is the $i$-th vertex of the $j$-th base, and $m$ is the number of eigenvectors.

In order to ensure the stability and consistency of the results, we apply a probability distribution to the coefficients $\pmb{\omega}_{3DMM}$ as a form of regularization:
\begin{equation}\label{eq3}
E_{3DMM-reg} = \sum_{i = 1}^{m}||\pmb{\omega}_{3DMM}^i / \pmb{\sigma}_i||^2_2,
 \end{equation}
where $\pmb{\sigma}_i$ is the relative eigenvalue of the eigenvector.

Hence, we can establish the complete cost function as follows:
\begin{equation}\label{eq4}
E_{3DMM} = E_{3DMM-plane} + \alpha_{3DMM} E_{3DMM-reg},
 \end{equation}
where $\alpha_{3DMM}$ represents the regularization parameter. The correspondence error evaluates the disparity between the input face and the estimated face, while the regularization term guarantees that the weights are consistent and not excessively distant from zero. By minimizing $E_{3DMM}$, we can achieve the ideal weights that minimize the deviation between the input face and the 3D facial model.

\subsection{Non-rigid ICP}

Non-rigid ICP is utilized to enhance the model by eliminating any rigid deformation. To ensure the consistency and uniformity of the model, we utilize a measurement technique proposed by Brian\cite{amberg2007}. We determine the rotation $\pmb{R}_i$ and translation $\pmb{t}_i$ of the $i$-th vertex.

The primary technique used to quantify the discrepancy between the distorted template and the target is through the distance between corresponding points. The cost function can be expressed as:
\begin{equation}\label{eq5}
E_{NICP-dis}(\pmb{X}) = \sum_{\pmb{v}_i\in \pmb{\mathrm{\nu}}} ||\pmb{\tau}_i - \pmb{X}_i \pmb{v}_i||_2^2,
 \end{equation}
where $\pmb{X}_i$ is constructed from the rotation $\pmb{R}_i$ and translation $\pmb{t}_i$ components, such that $\pmb{X}_i = [\pmb{R}_i | \pmb{t}_i]$. Also $\pmb{v}_i$ represents the homogeneous coordinates of the $i$-th vertex, while $\pmb{\tau}_i$ is a set of target points that are related to the vertex.

To ensure rigidity, we penalize the difference in transformation between neighboring vertices using the Frobenius norm $||\cdot||_F^2$. Specifically, we define the rigidity term as follows:
\begin{equation}\label{eq7}
E_{NICP-ARAP}(\pmb{X}) = \sum_{\{i, j\}\in\pmb{\varepsilon}}||(\pmb{X}_i - \pmb{X}_j)\pmb{G}||^2_F,
 \end{equation}
where $\pmb{G} = diag(1, 1, 1, \gamma)$, $\gamma$ is used to balance the rotation and translation. $\pmb{\varepsilon}$ is the set of two-point-sequential pairs of edges. Based on As-Rigid-As-Possible surface deformation (ARAP)\cite{sorkine2007}, we refine the smoothing item with the vertices' position as follows:
\begin{equation}\label{eq9}
\sum_{j \in \mathcal{N}(i)}\pmb{\omega}_{ARAP}^{ij}(\pmb{p}'_i - \pmb{p}'_j) = \sum_{j \in \mathcal{N}(i)}\frac{\pmb{\omega}_{ARAP}^{ij}}{2}(\pmb{R}_i + \pmb{R}_j)(\pmb{p}_i - \pmb{p}_j).
 \end{equation}

we refine the smoothing item with the vertices position,
\begin{equation}\label{eq10}
E_{NICP-ARAP}(\pmb{X}) = \sum_{\{i, j\}\in\pmb{\varepsilon}}\pmb{\omega}_{ARAP}^{ij}||(\pmb{X}_i - \pmb{X}_j)\pmb{G}\frac{(\pmb{v}_i+ \pmb{v}_j)}{2}||^2_2.
\end{equation}

Compared to the original form, this form eliminates the normalization process, making the calculation more convenient.

To regular the rotation and translation, define
\begin{equation}\label{eq11}
E_{NICP-reg}(\pmb{X}) = \sum_{i} ||\pmb{X}_i\pmb{X}_i^T - \pmb{I}||^2_F.
 \end{equation}

The full cost function is a weighted sum of these terms
\begin{align}
E_{NICP}(\pmb{X}) =& E_{NICP-dis}(\pmb{X}) \notag \\
+& \alpha_{NARAP} E_{NICP-ARAP}(\pmb{X}) \notag \\
+& \alpha_{Nreg} E_{NICP-reg}(\pmb{X}),\label{eq12}
\end{align}
which consists of three terms: the correspondence term $E_{NICP-dis}(\pmb{X})$, the regularization term $E_{NICP-reg}(\pmb{X})$, and the smoothing term $E_{NICP-ARAP}(\pmb{X})$. The correspondence term measures the distance between the deformed template and the target through the distance between the corresponding points. The regularization term ensures that the rotation and translation of the vertices are close to the identity matrix. The smoothing term maintains the rigidity of the model by penalizing the difference of the transformation of neighboring vertices. The weights $\alpha_{NARAP}$ and $\alpha_{Nreg}$ are used to balance the influence of the smoothing term and the regularization term. In addition, if necessary, a landmark term can be added to further improve the accuracy of the correspondence points.

\section{Drive}

The key issue for real-time facial animation is to solve the equation for $\pmb{\omega}_{bs}^t$:
\begin{equation}\label{eq13}
\pmb{b}_t = \pmb{B\omega}_{bs}^t + \pmb{b}_0,
 \end{equation}
where $\pmb{b}_t$ is a vector consisting of an ordered arrangement of points on the mesh corresponding to the $t$-th frame, $\pmb{b}_0$ is a vector consisting of an ordered arrangement of points on the neutral expression mesh, and $\pmb{B}$ is a matrix consisting of residual sequences between other blendshapes and the neutral. In this case, 51 blendshapes (excluding the tongue out) are taken according to the ARKit standard.

\subsection{Weights Calculation}
The cost function of weight $\pmb{\omega}_{bs}$ is given by:
\begin{align}
		E_{RT}(\pmb{\omega}_{bs}^t) =& E_{RT-dense} (\pmb{\omega}_{bs}^t) \notag \\
        +& \alpha_{Rmarks} E_{RT-marks} (\pmb{\omega}_{bs}^t)  \notag \\
		+& \alpha_{Rsmooth} E_{RT-smooth}(\pmb{\omega}_{bs}^t) \notag \\
        +& \alpha_{Rreg} E_{RT-reg}(\pmb{\omega}_{bs}^t), \label{eq14}
\end{align}
where
\begin{align}
		E_{RT-dense}(\pmb{\omega}_{bs}^t) =& ||\pmb{S}_{dense}(\pmb{B}\pmb{\omega}_{bs}^t+ \pmb{b}_0) - \pmb{m}_{dense}||_2^2, \label{eq15} \\
		E_{RT-marks}(\pmb{\omega}_{bs}^t) =& ||\pmb{S}_{marks}(\pmb{B}\pmb{\omega}_{bs}^t + \pmb{b}_0) - \pmb{m}_{marks}||_2^2, \label{eq16} \\
		E_{RT-smooth}(\pmb{\omega}_{bs}^t) =& ||\pmb{\omega}_{bs}^t - 2\pmb{\omega}_{bs}^{t-1} + \pmb{\omega}_{bs}^{t - 2}||^2_2, \label{eq17} \\
		E_{RT-reg}(\pmb{\omega}_{bs}^t) =& ||\pmb{\omega}_{bs}^t||_2^2. \label{eq18}
\end{align}

The matrices $\pmb{S}_{dense}$ and $\pmb{S}_{marks}$ represent the dense select matrix and the landmarks select matrix, respectively. The dense select matrix $\pmb{S}_{dense}$ is a binary matrix that has the same number of rows as the number of vertices in the 3D mesh model. It selects the vertices that are used to create the dense correspondences between the face model and the input mesh. On the other hand, the landmarks select matrix $\pmb{S}_{marks}$ is also a binary matrix but has a smaller number of rows than $\pmb{S}_{dense}$. It selects the vertices that correspond to the landmarks on the face, which are used as reference points for the 3D model fitting process.

The purpose of the regularization term is to make the fitted result as close as possible to the neutral expression without excessive deformation.
\begin{itemize}
\item To improve the solving speed and accuracy, the face area is divided into pieces for parallel computation.
\item If source is aligned to target for computation, the residual matrix $\pmb{B}$ needs to be recalculated each frame. To simplify the computation process, the inverse of transformation is applied to the target for solving.
\end{itemize}

\subsection{Weights Filter}

To improve the dependency between blendshapes, we propose an optimization model based on residual neural networks. The dataset used to train the model is composed of 20,000 frames. Each sample consists of the ground truth of the previous $n$ frames and the solving result of the current frame as input, and the ground truth of the current frame as output. By using this dataset, we aim to improve the consistency and realism of the generated facial animations.

The proposed model is based on residual neural networks, which are a type of neural network architecture designed to address the vanishing gradient problem in deep neural networks. The model takes advantage of the residual connections between layers, allowing the network to learn more complex representations and achieve better performance.

By training the model on the dataset, we can filter and regularize the current frame, improving the dependency between blendshapes and reducing discontinuities in the animation. This approach provides a more realistic and consistent facial animation, with better overall quality.

\subsection{Eye Gaze}

After performing facial reconstruction, it is possible to obtain the position of the eyeballs. During real-time driving, registration can be performed to determine the orientation of the head and align the 3D model with the 2D image. By detecting facial landmarks, the 2D and 3D information of the pupils can be aligned, allowing the calculation of the rotation angle of the eyeballs. This information can then be used to animate the face more realistically and expressively.

As the localization of facial landmarks may have some degree of error, relying solely on the center position of the eyeball to determine its current state can lead to inaccuracies. To address this, we use the contour of the pupil as an additional reference to assist in the calculation. The cost function for the rotation $\pmb{R}_{eye}$ of the eyeballs can be obtained using the projection matrix $\pmb{P}$. Since $\pmb{P}$ maps 3D points in the eye model to their 2D counterparts in the camera image, it can be obtained from the camera as it is fixed. So
\begin{align}
E_{eye-dis}(\pmb{R}_{eye}) &= ||\pmb{PRc}^{3D}_{model} - \pmb{c}^{2D}_{image}||_2^2  \notag \\
&+ \alpha_{Edis}||1 - \frac{S_{overlap}}{S_{pupil}}  ||_2^2, \label{eq13}
\end{align}
where the variable $S_{overlap}$ represents the area of overlap between the projection of the pupil onto the 2D image and the detected pupil in the 2D image. The vector $\pmb{c}^{3D}_{model}$ represents the center of the iris in the 3D model, while $\pmb{c}_{image}^{2D}$ represents the center of the detected iris in the 2D image.

To increase the continuity of the eyeball movement over time, we add a smoothing term,
\begin{equation}\label{eq13}
E_{eye-smooth}(\pmb{R}_{eye}) = ||\pmb{R}_{eye} - 2\pmb{R}_{eye}^{t-1} + \pmb{R}_{eye}^{t}||_F^2.
 \end{equation}
To minimize the full cost function,
\begin{align}
E_{eye}(\pmb{R}) &= E_{eye-dis}(\pmb{R}_{eye})  \notag \\
&+ \alpha_{Esmooth} E_{eye-smooth}(\pmb{R}_{eye}),\label{eq13}
\end{align}
which includes both the cost function for pupil contour alignment and the smoothing term, we can obtain a more natural and vivid rotation of the eyeball. Here, $\alpha_{eye}$ is a weight coefficient that balances the influence of the two terms.

\section{ Experimental Results}

We implemented our system on a PC with an Intel Core i7 (3.5GHz) CPU and a 3D camera (recording 640 x 480 images at 30 fps). The parameters are show in Table. \ref{table1}.

In this study, we fixed the camera and rotated the head to capture facial expressions from various angles. These point clouds were then processed through a modeling and fusion pipeline to generate a series of personalized blendshapes for each user. Non-rigid ICP registration was applied to the fused results to further refine the blendshapes.

\begin{figure*}[htbp]
	\centering
	\subfigure[Individual 1]{
		\begin{minipage}{27pc}
            \includegraphics[width=\textwidth]{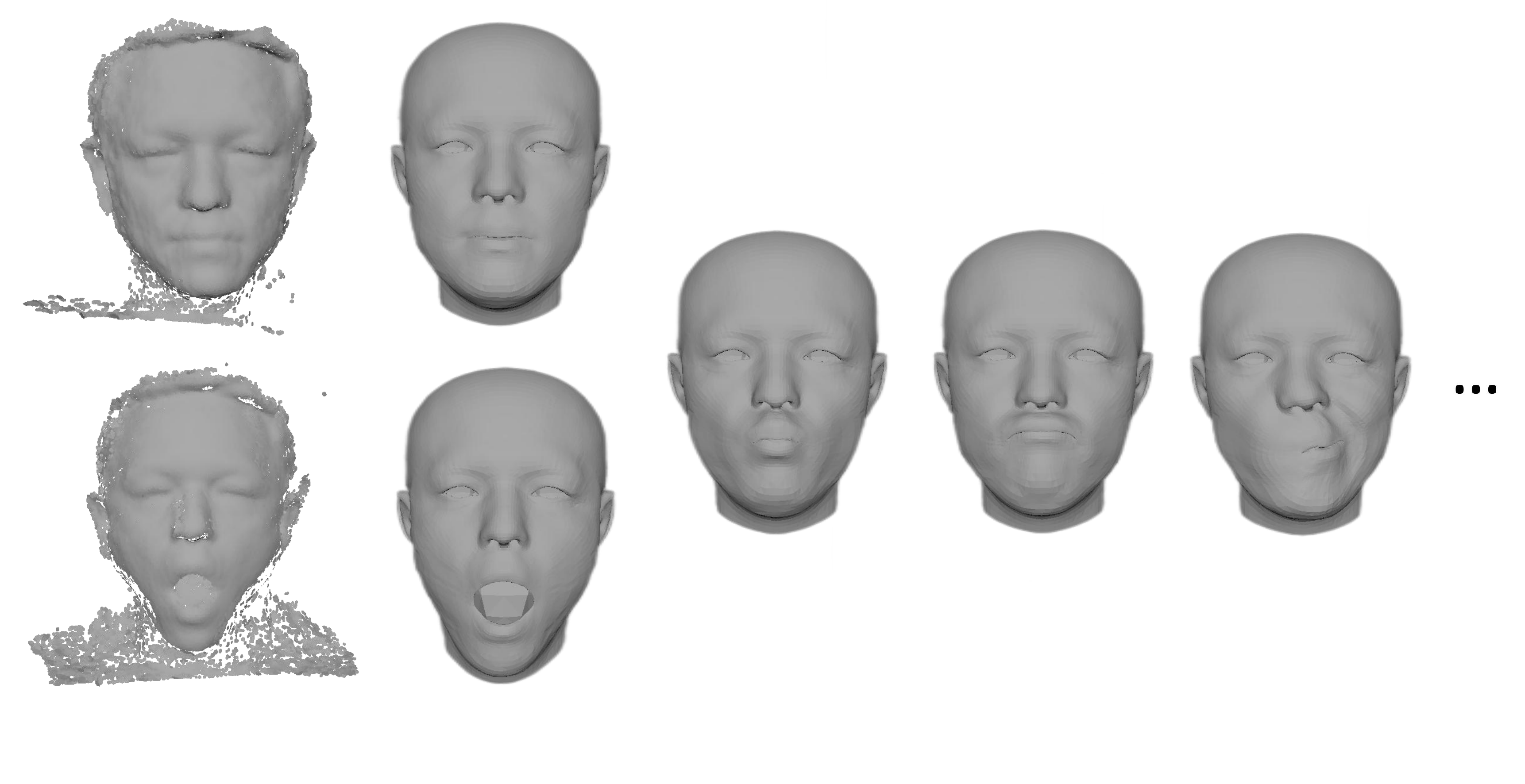} \\
		\end{minipage}
	}
	\subfigure[Individual 2]{
		\begin{minipage}{27pc}
            \includegraphics[width=\textwidth]{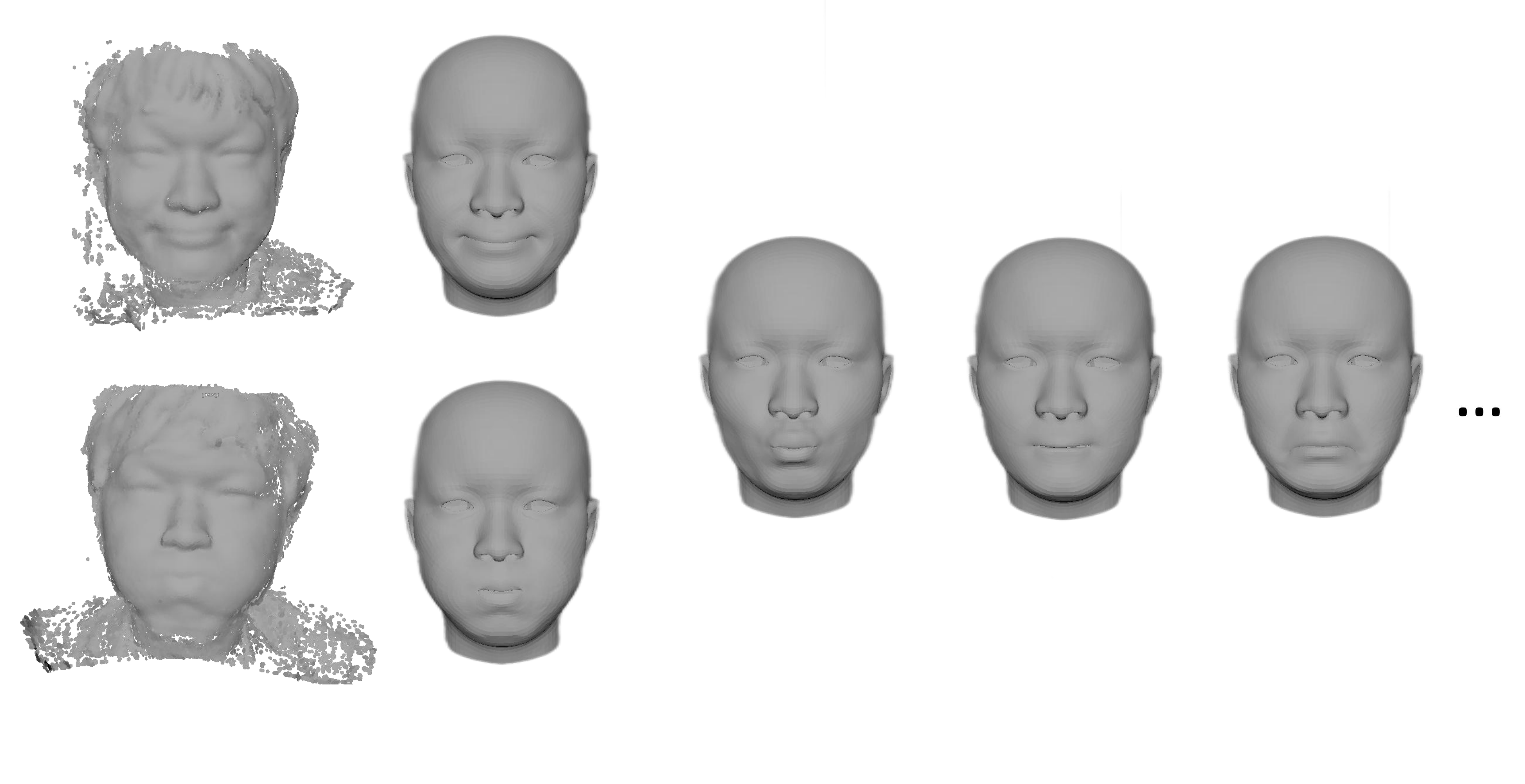} \\
		\end{minipage}
	}
	\subfigure[Individual 3]{
		\begin{minipage}{27pc}
            \includegraphics[width=\textwidth]{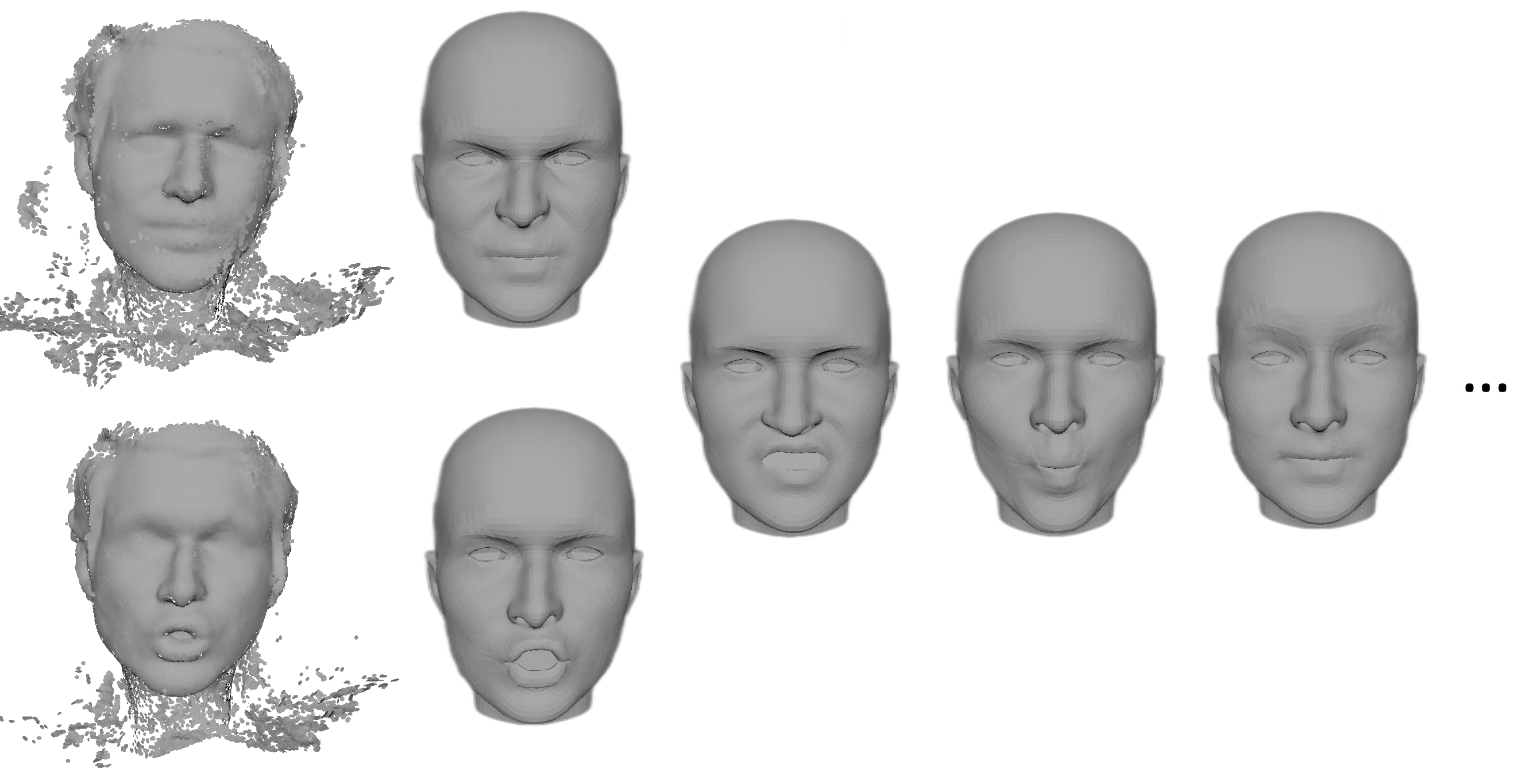} \\
		\end{minipage}
	}
	\label{fig:3}
	\caption{The first column is the fused point cloud, and the second column is the model after non-rigid registration. The following ones are other blendshapes.}

\end{figure*}

Figure. \ref{fig:3} shows the fusion and registration results for different individuals and poses. The blended mesh effectively capture the unique characteristics of each user's facial expressions, producing a set of blendshapes that accurately represent their individual features. The Non-rigid ICP registration process further improves the accuracy of the blendshapes, resulting in a more realistic and natural appearance.

The real-time driving effect will be demonstrated in subsequent videos.

The input point clouds were captured using a depth camera, and the output results were generated through our real-time driving algorithm. The algorithm is based on a combination of facial tracking and animation techniques, which enable us to map the user's facial movements onto a 3D avatar in real-time.

As can be seen in the figure, our algorithm is capable of accurately capturing and reproducing the user's facial expressions, including subtle nuances and movements. The resulting 3D avatar closely mimics the user's facial movements, creating a natural and expressive appearance.

These results demonstrate the potential of our approach for real-time facial animation in various applications, such as video conferencing, gaming, and virtual reality. By enabling real-time interaction and communication with lifelike avatars, our approach can enhance the user's sense of presence and immersion in virtual environments

\begin{table}[h]
\centering
\caption{the parameters}
\begin{tabular}{c|c}
\toprule
$\alpha_{3DMM}$ & 0.1  \\
$\alpha_{NARAP}$ & 20   \\
$\alpha_{Nreg}$ & 1 \\
$\alpha_{Rmarks}$ & 0.5   \\
$\alpha_{Rsmooth}$ & 0.2 \\
$\alpha_{Rreg}$ & 0.5   \\
$\alpha_{Edis}$ & 1  \\
$\alpha_{Esmooth}$ & 0.5  \\
\bottomrule
\end{tabular}
\label{table1}
\end{table}

\section{Conclusion}

Real-time facial animation is a complex and challenging task that requires accurate and efficient processing of various facial features. In this text, we have explored the methods and techniques used in real-time facial animation, such as weight calculation, weight filtering, and eye gaze. These techniques use mathematical models and algorithms to track facial landmarks and eyeball movements accurately, resulting in more natural and expressive facial animations.

We have successfully implemented a real-time driving system based on 52 blendshapes. Our algorithm is compatible with a wider range of blendshapes, allowing us to achieve even higher accuracy in facial capture.

\bibliographystyle{unsrtnat}
\bibliography{reference}

\end{document}